\documentstyle[11pt,epsfig]{article}

\def\barr{\begin{equation} \begin{array}{*{4}{c}} }
\def\earr{\end{array} \end{equation}}
\def\be{\begin{equation}}
\def\ee{\end{equation}}
\def\del{\partial}
\def\g{\gamma}
\def\tg{\tilde \gamma}
\def\a{\alpha}
\def\o{\omega}

\begin{document}

\title{A model for ripple instabilities in granular media}

\author{Orestis Terzidis, Philippe Claudin and Jean-Philippe Bouchaud\\
Service de Physique de l'Etat Condens\'e, C.E. Saclay\\
Orme des Merisiers, 91191 Gif-sur-Yvette Cedex, France}
\date{\today}
\maketitle

\begin{abstract}
We extend the model of surface granular flow proposed in \cite{bcre} to
account for the effect of an external `wind', which acts as to dislodge particles from the static bed, such that a stationary state of flowing grains 
is reached. We discuss in detail how this mechanism can be described in a
phenomenological way, and show that a flat bed is linearly unstable against ripple formation in a certain region of parameter space. We focus in particular on the (realistic) case where the migration velocity of the instability is much smaller than the grains' velocity. In this limit, the full dispersion relation can be established. We find that the critical wave vector is of the order of the saltation length. We provide an intuitive interpretation of the instability.
\end{abstract}

\section{Motivation}

Common observations suggest that flat sand surfaces can become unstable when
subjected to moving air or water. After some time regular patterns appear, as
can be observed
on desert dunes, underwater sand, `dry' snow, etc. These patterns resemble
surface waves; however their physics is completely different since in the case
of sand there is no surface tension. Following Bagnold (\cite{bag}, chap. 11)
these patterns can be classified into  ripples, ridges and dunes. The
repetition distance of ridges varies with time, whereas ripples exhibit a
stationary wavelength after some transient. Early qualitative arguments by
Bagnold \cite{bag} suggested that the ripple wavelength $\lambda$ is related
to the typical path length of the blown grains, called the `saltation length'
$\xi$. A more quantitative `two-species' model was proposed by Anderson
\cite{and}, which describes the coupling between the moving grains and the
static bed. Such a model predicts that a flat surface is unstable for all
wavelengths, with a faster growing mode indeed comparable to the typical
jump length of the grains. However, this model is incomplete: while the
 dynamics of the static bed is treated exactly, the description of the moving phase is highly simplified. Alternatively, there are also several numerical
models for ripple formation \cite{lawe}. In this paper, we extend Anderson's theoretical model of 
ripple formation,
by adapting the phenomenological equations for surface flow
introduced in \cite{bcre} in the context of avalanches, and further discussed 
in \cite{dg,boutdg,stanley,cargese,triangular}.

It is worth recalling, after Bagnold \cite{bag}, some basic facts about the
motion of the grains and the formation of these patterns: (i) There are two
qualitatively distinct transport mechanisms for the grains, saltation and
surface creep \footnote{To which one should also add `suspension', corresponding to very small grains flying high in the air.}. The trajectories of grains in saltation is determined by the
velocity profile of the wind, the air friction limiting the grain velocity and
by the initial energy of the grain when first expelled from the sand bed. One 
of the characteristic features of the path is the flat angle of incidence
which varies between $10^{\rm o}$ and $15^{\rm o}$. (ii) The saltation has two
effects on the surface: it either rebounds and/or ejects grains leading to a
new saltation or it produces surface creep. There is however no sharp boundary
between these two processes, since the energy of the ejected grains varies 
continuously. Both saltation and creep lead to a net flow of grains in the
direction of the wind.  (iii) The time scale of ripple formation is much
larger than that of saltation. (iv) The migration velocity of the ripples is
much smaller than a mean transport velocity (averaged over saltation and
creep).

\section{A `two-species' model with wind}

The phenomenological approach we consider in the following is based on the
observation that two different species of grains enter the problem: moving
grains and grains at rest. We will not distinguish between grains in saltation
and creep, but introduce an appropriately averaged quantity describing grains
that are convected by either of the two mechanisms, which we call the moving
grain density $R(x,t)$ \footnote{In principle, one should consider a density
$R(x,v,t)$ which depends on the instantaneous velocity of the grains. 
$R(x,t)$ is the average of $R(x,v,t)$ over all velocities.}, where $x$ is the
coordinate in the direction of the wind and $t$ the time. (We will assume that
the problem is translationally invariant in the direction transverse to the
wind; see \cite{triangular} for an extension to two dimensions). The grains at
rest contribute to the local height $h(x,t)$ of the static bed. The dynamical 
equations for $R$ and $h$ read, in the hydrodynamical (long wavelength) limit:
\begin{eqnarray}
\del_t R&=&-V\del_xR+D_1\del_x^2R+\Gamma[R,h]\nonumber\\
\del_t h&=&-\Gamma[R,h]
\end{eqnarray}
where $V$ and $D_1$ are the average velocity of the grains and the dispersion
constant, related to the fact that grains do not all move with the same
velocity \footnote{These terms can be understood, more generally, as the long-wavelength limit of a more general non-local convection 
term of the kind $\int
K(x-x')R(x',t)\,dx'$.}. $\Gamma$ describes the rate with which a grain at rest
is converted into a moving grain (or vice versa) and depends both on $R$ and
on the local surface profile. For simplicity we have defined $R$ to have the
same dimension as $h$, and it can be thought of as the width of grains which
has been removed from the static bed. Correspondingly, $\Gamma$ has the
dimension of a velocity.  
The construction of $\Gamma$ is based on phenomenological arguments
\cite{bcre}, and encodes different physical processes: 

$\bullet$ Due to the presence of wind, grains can be `spontaneously' ejected
from the surface, even in the absence of already moving grains. The rate at
which this occurs depends on the local wind velocity (or rather velocity
gradient at the surface). Since the wind velocity tends to be larger when the
local slope is facing the wind, we write:
\be
\Gamma_{sp} = \a_0+\a_1\del_xh-\a_2\del_x^2h
\ee
where the coefficients $\a$ are positive or zero. We have also included the second
derivative contribution with a minus sign, since grains are harder to dislodge
in troughs than at the top of a crest. Note  that all these coefficients
are expected 
to depend on the external wind velocity. In particular, as shown by Bagnold
himself, the coefficient $\a_0$ is only non-zero above a certain critical wind
velocity, noted $V^*_{\rm{fluid}}$. 

$\bullet$ When hitting the ground, a moving grain can either be captured or
transfer a part of its kinetic energy to other static grains and provide new
moving particles. The rate at which both these process occur is proportional
to $R$ (at least for small enough $R$ -- see below), and also depends 
on the wind velocity and on the local slope. For example, flying grains have a
larger probability to land on a surface facing the wind, rather than in the
wind shadow. This suggests to write the {\it stimulated} conversion rate as:
\be
\Gamma_{st}=-R[\g_0+\g_1\del_x h+\g_2\del_x^2h]
\ee
The sign of $\g_0$ depends on the strength of the wind; for small wind
velocity, one expects capture to be more important than emission, and thus
that $\g_0 > 0$. As again shown by Bagnold, a localized source of moving
grains 
tends to die away when the wind velocity is less than a certain
$V^*_{\rm{impact}}<V^*_{\rm{fluid}}$, whereas a steady saltation is found for
larger velocities, suggesting that $\g_0<0$ for $V>V^*_{\rm{impact}}$. 
In this case,
however, it is easy to see that $R$ increases 
exponentially, and that higher order terms are needed to describe the
stationary situation. One can think of several non-linear effects: for
example, collision between flying grains leads to dissipation and hence to a
poorer
efficiency of the impacts on the static bed. Also, the presence of a layer of
moving grains screens the hydrodynamical flow, which in turn reduces the
energy transfer between the wind and the saltating grains. These effects can 
be described by a term proportionnal to $-\beta R^2$ in $\Gamma_{st}$
\footnote{In principle, the dependence of $V$ on $R$ should also be taken into
account. We do not consider this here, since this does not affect the linear
instability analysis.}. If trapping dominates (as is the case 
for under water ripples) one expects $\g_1 >0$ because more grains fall on the
slope facing the convective flow. For the same reason, if stimulated emission dominates,
as is the case for wind blown sand, one expects that $\g_1 < 0$. Finally,
$\g_2$ is positive since, again, 
grains are easier to dislodge at the top of a bump.

The total conversion rate $\Gamma$ is obtained as the sum of $\Gamma_{sp}$
and $\Gamma_{st}$, while the model proposed in \cite{bcre} did not contain 
the wind induced contribution proportionnal to $\a$, nor the non-linear term.
The equation for $h$ thus reads:
\be
\del_t h = (R \g_0 - \a_0) + \beta R^2 + (R \g_1 - \a_1) \del_x h + (R \g_2 +
\a_2)\del_x^2h 
\ee
The gradient term can be interpreted as a translation of the surface profile
with time, at velocity $\a_1 - R\g_1$. The direct action of the wind ($\a_1$) is indeed
to erode grains from the windward slope of a bump and transport them in the
direction of the wind. The other contribution ($R\g_1$), however, moves the bumps 
`backwards' since grains are effectively deposited on the windward slope,
contributing to a translation of the bump against the wind. (A similar
discussion can be found in \cite{bcre,cargese,triangular}.)

Note that the above set of equations is non-linear, so that non-trivial
dynamics is expected. Some essential features of the model can be investigated
by linearizing the system in the vicinity of the situation where the surface
is flat ($h_0=0$). The moving grain density is then equal to (see Fig 1):
\be
R_0= \frac{1}{2 \beta} \left[-\g_0 + \sqrt{\g_0^2 + 4 \alpha_0 \beta}\right].
\ee

\begin{figure}[t] 
\begin{center}
\psfig{figure=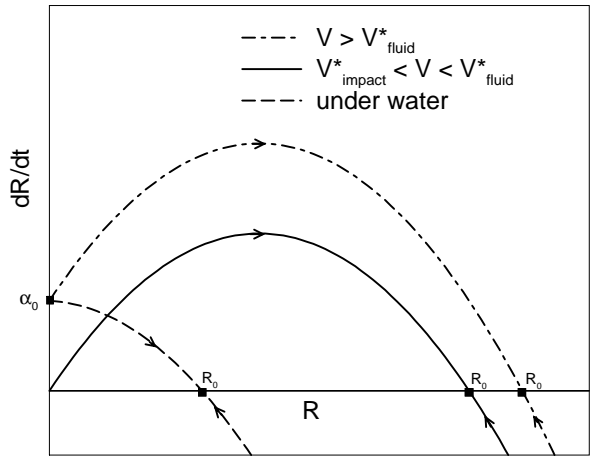,height=8cm}
\caption[]{\small{Stability diagram, showing $dR/dt$ as a function of $R$ in an 
homogeneous situation. The case $\g_0 < 0$ corresponds to blown sand with 
$V > V^*_{\rm{impact}}$, where stimulated emission is very efficient, and where $R_0$ can be non-zero even if
$\a_0 = 0$ (i.e. when $V < V^*_{\rm{fluid}}$). The situation where capture dominates ($\g_0 > 0$) is probably relevant for sand under water.}}
\end{center}
\end{figure}

\section{Stability analysis}

We will perform a stability analysis, i.e. investigate whether a small
perturbation is amplified or dies out with time. Therefore we consider
$R=R_0+{\bar R}$, $h=h_0+{\bar h}$ and neglect second order terms of the kind
${\bar R}{\bar h}$, ${\bar R}^2$ and ${\bar h}^2$. For simplicity of notation
we drop the bars; the linearized equations then read
\begin{eqnarray}\label{lin}
\del_tR&=&-\tg_0R-V\del_xR+D_1\del_x^2R+W\del_xh-D_2\del_x^2h+...\nonumber\\
\del_t h&=&\tg_0R-W\del_xh+D_2\del_x^2h+...
\end{eqnarray}
with an effective velocity $W=\a_1-R_0\g_1$ and an effective diffusion
constant $D_2=R_0\g_2+\a_2 > 0$. $\tg_0$
is equal to $\g_0 + 2 \beta R_0$ and is thus always {\it positive}. A Fourier
analysis of the linearized equations leads to
\be
\left(
\begin{array}{cc}
-i\o-\tg_0-ikV-k^2D_1&ikW+k^2D_2\\
\tg_0&-i\o-ikW-k^2D_2\\
\end{array}
\right)
\left(
\begin{array}{cc}
\widetilde{R}\\
\widetilde{h}\\
\end{array}
\right)=0
\ee
where the tilde denotes the Fourier transforms. This system has a non-trivial
solution if the determinant of the above matrix is zero, leading to the
relation
\be\label{wq}
\o^2+\o(a+ib)+(c+id)=0.
\ee
The coefficients read
\begin{eqnarray}\label{abcd}
a&=&(V+W)k\nonumber\\
b&=&-\left[ \tg_0+(D_1+D_2)k^2\right]\nonumber\\
c&=&VWk^2-D_1D_2k^4\nonumber\\
d&=&-(D_1W+D_2V)k^3;
\end{eqnarray}
they are functions of the wave vector $k$ and of the system's parameters
($V$, $W$, $D_1$, $D_2$, $\tg_0$). 

Equation (\ref{wq}) establishes a dispersion relation $\o(k)$ with two
branches corresponding to the two solutions of the quadratic equation, where
$\o$ has to be considered as a complex variable. (Writing down the
corresponding equations for the real and the imaginary part of $\o$ leads to
quartic equations.) In the context of a stability analysis we are interested
in the imaginary part of $\o(k)$: as long as it is positive $e^{i\o t}$ will
decay exponentially, while a negative imaginary part does lead to an
instability. This imaginary part is given by:
\begin{eqnarray}
&2&{\rm Im}(\o_{\pm})=\nonumber\\
&-&b\pm{1\over\sqrt{2}}\left[
-(a^2-b^2-4c)+\left[
(a^2-b^2-4c)^2+(2ab-4d)^2
\right]^{1/2}
\right]^{1/2}
\end{eqnarray}
which is a function of $k$. A critical wave vector $k^*$ can be defined such
that Im($\o$) is exactly zero, which leads to $
d^2-abd+b^2c=0$. Inserting the explicit expressions (\ref{abcd}), one finds a
cubic equation for $k^{*2}$. Whenever this equation admits a positive
solution, there will be a finite band of wave vectors $[0,k^*]$ which are
unstable (see Figure 2).
\begin{figure}[t] 
\begin{center}
\psfig{figure=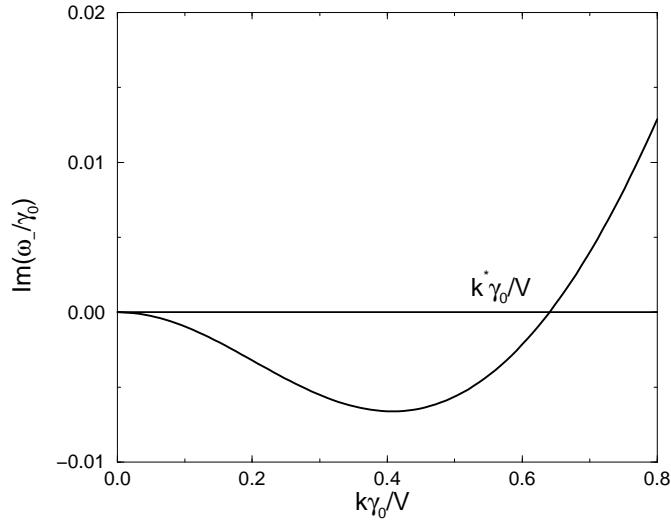,height=8cm}
\caption[]{\small{Rescaled damping rate as a function of  the rescaled wave vector.
The plot shows data for $\eta=0.1$, $\tg_0 D_1=V^2$ and $D_2/D_1=0.1$}}
\end{center}
\end{figure}

It is instructive to study the asymptotic behaviour of the functions ${\rm
Im}(\o_-)$. One finds:
\be
{\rm Im}(\o_-)=\cases{
-{\eta v^2\over\tg_0}\; k^2+... & for $k\ll \tg_0/V$\cr 
D_2  k^2+... & for $k\gg \tg_0/V$\cr}.
\ee
with $\eta=W/V$. The transport velocity $V$ (by convention) and the diffusion
constants are positive; the main control parameter remaining is the relative
migration velocity $\eta$. One sees that for $0<\eta<1$ there is indeed a band
of instable wave vectors (one can see from the asymptotic solutions that the
sign changes for large $k$'s). The situation where $\eta<0$ is stable and $\eta>1$ (i.e. a
bump moving faster than the transport velocity) does not seem physical. The
second branch Im$(\o_+)$ is always positive and is thus of no importance for
our stability considerations.

Following the intuition that the ripples move much more slowly than the grains
are transported, we will assume in the sequel that $0<\eta\ll 1$, which we
attribute to the fact that the $\alpha$ coefficients are small compared to
$V$.
Since $D_2 \propto \a$, this suggests that the diffusion constants $D_1$ and
$D_2$ are in the same ratio, so we write: $D_2=\delta\eta D_1$, where $\delta$
is of the order of one. These assumptions make it possible to simplify the
algebra and to find the solution:
\be\label{wapp}
{\rm Im}(\o_-)=\eta\;
{k^2\left[-\tg_0 V^2+D_1\delta(\tg_0 D_1+V^2)k^2+\delta D_1^3k^4\right]
\over \tg_0^2+k^2(2\tg_0 D_1+V^2) +k^4D_1^2}+o(\eta^2).
\ee
which is plotted in Figure 2. As we discuss now, three relevant facts can be
verified with this formula: (i) the critical wave vector is of the order of
the inverse mean saltation length, (ii) the ripple velocity is of the order of
$\eta V$ and (iii) the time scale of ripple formation is much larger than the
saltation time scale. 

Let us first give some arguments for (i). Since the saltation trajectories
result from some random initial vertical velocity of the grains, the saltation
lengths will also be random, with both short jumps (actually corresponding to
creep) and long jumps. It is reasonable to assume that the width of the
saltation length distribution is of the same order as its mean $\xi$ (a
similar assumption is discussed in
\cite{and}). In this situation, the `P\'eclet' number defined as
${\rm{Pe}}=V \xi/D_1$ is of order one: convective and diffusive effects are
of the same order of magnitude. In the case where the jump length distribution
is sharply peaked around $\xi$, one would rather have ${\rm{Pe}} \gg 1$.

Defining $\xi = V\tau$, where $\tau$ is
the typical saltation time, one finds that the the zero of (\ref{wapp}) is
located at:
\be
k_*^2={{\rm Pe}\over 2\xi^2}\;
\left\{\sqrt{(\tg_0\tau+{\rm{Pe}})^2+4
\tg_0\tau/\delta}-({\rm{Pe}}+\tg_0\tau)\right\}
\ee 
Since $\tg_0$ is the (renormalized) rate of sticking, it is reasonnable to
assume that $\tg_0 \tau \sim 1$, thereby
confirming that $k^* \sim \xi^{-1}$ for ${\rm{Pe}} \sim 1$. On the other
hand, for weakly dissipative collisions (hard grains) one expects that $\tg_0 \tau \ll
1$, leading to larger unstable wavelengths $\sim \xi\sqrt{\delta/\tg_0 \tau}$.

The ripple velocity is given by the corresponding dispersion relation, i.e.
the real part of $\o(k)$. One finds:
\be
2{\rm Re}(\omega_-)=\eta\;
{k^3V\left[V^2+\tg_0D_1(1+\delta)+D_1^2k^2\right]\over
\tg_0^2+k^2(2\tg_0 D_1+V^2) +k^4D_1^2}+o(\eta^2).
\ee
The formula shows that for $k \sim k^*$, both phase and group velocities
are indeed of the order of $\eta V$, establishing thus (ii).

Finally knowing the fastest growing wave vector, one finds that the ripple
formation time $t_{\rm ripple}$ (determined by the depth of the minimum in
figure 2) is a factor $1/\eta$ larger than say $1/\tg_0$ or $\tau$, i.e. that
ripple formation occurs on much slower time scales than any microscopic
process. The ratio of formation time and microscopic time scales should indeed
be roughly the same as that between migration and convection velocity (iii).

\section{Physical discussion and open questions}

Let us finally give an intuitive interpretation of the instability. Imagine a
flat surface with a finite number of moving grains above it (i.e. the
stationary solution). Now imagine a small perturbation of this situation, say
a small hump. The term $\del_t R \sim \del_x h$ in the linearized equations
(\ref{lin}) increases locally the concentration of the moving grains thus
producing a `cloud' at the windward side of the hump. This cloud is convected
with the velocity $V$and after a time unit of $1/\tg_0$ the cloud has moved a distance $\xi$ where
the cloud starts to `rain' (i.e. moving grains are converted into grains at
rest). If the position of the hump has in the same time moved in the same direction, its height will increase, leading to an instability. (Conversely, if  the bump moves backward -- i.e. if $W <0$ -- the `rain' will rather 
fill the hole and smear out the bump). The presence of the
diffusive processes counterbalances the amplification for small distances and
some optimum wavelength of the order of $\xi$ (corresponding with the minimum
in figure 2) becomes visible.

Summarizing, we have thus shown that equations (1), which are
phenomenological, but motivated by clear physical 
processes, indeed show an instability which is consistent with some essential
features of ripple formation. It is worth noting that our analysis, which
concentrated on the linearized system in the vicinity of the stationary
solution, is universal in the sense that a whole class of models behaves in an
analogous way (with some possible redefinition of the coefficients). For
example, a non-linear dependence of the velocity $V$ on $R$ does not
modify the above analysis, up to a redefinition of $V$. Note
also that all phenomenological coefficients are, at least in principle,
measurable in situations independent from ripple formation (since they are
diffusion constants, convection velocities, deposition rates etc.). In this
sense it should be possible to check experimentally for the consistency of the
above description. 

Our conclusions are very similar to those reached by Anderson \cite{and}, on
the basis
of a simplified model where the flowing phase (what we have called $R$
above) is assumed to
be in equilibrium from the outset, and where a rather arbitrary distinction 
is made between
`saltating grains' which are never captured by the bed, and `reptating' grains
which are captured after exactly one jump. Correspondingly, 
the structure of the
dispersion
relations differ in the two approaches. Furthermore, it is difficult to extend
Anderson's model beyond the linear instability analysis while our model, in
principle, can account 
for non-linear effects \cite{usinprep}.

Finally, there are several open questions which we would like to mention and
leave for future work: (i) Can one establish some precise relations between
the `microscopic' coefficients (like wind velocity, polydispersity, elasticity
etc.) and the phenomenological parameters? (ii) How are the above results
modified if one considers two spatial dimensions? Is there an instability
corresponding to the transverse wavelike shape of the ripples known from field
observation? (iii) What is the ripple shape and height predicted from a non-linear analysis of the equations? (iv) Is it important to consider a non-local
convection term, rather than the hydrodynamical form written in (1)? The
question arises since the relevant wavelength is precisely of the same order
as (and not much larger than) the jump length $\xi$.

\vskip 1cm

{\sc acknowledgements:} We want to thank A. Valance and F. Rioual for very useful discussions. The Coll\`ege de France lectures of P.G. de Gennes on blown sand have also been of great help in our understanding of the subject. OT thanks the French Foreign Office and the CROUS de Versailles for a post-doctoral grant.


\begin{thebibliography}{99}

\bibitem{bcre} J.P. Bouchaud, M.E. Cates, R. Prakash, S.F. Edwards: J. Phys.
France {\bf 4} (1994) 1383, Phys. Rev. Lett. {\bf 74}, (1995) 1982

\bibitem{bag} R. A. Bagnold: The physics of blown sand and desert dunes
(1941), Reprinted by Chapman and Hall (1981)

\bibitem{and} R.S. Anderson: Sedimentology {\bf 34} (1987), 943-956, Earth Science
Reviews {\bf 29} (1990), 77-96

\bibitem{dg} P.G. de Gennes, Comptes Rendus Acad\'emie des Sciences, {\bf
321} II (1995) 501, Lecture Notes, Varenna Summer School on Complex
Systems, July 1996.

\bibitem{boutdg} T. Boutreux, P.G. de Gennes, J. Phys. I France, {\bf 6}
(1996) 1295.

\bibitem{stanley} H. A. Makse, S. Havlin, P. R. King, H. E. Stanley,
Nature (London) {\bf 386 } (1997) 379, H. A. Makse, P. Cizeau, H. E.
Stanley, Phys. Rev. Lett. {\bf 78} (1997) 3298.

\bibitem{cargese} J.P. Bouchaud, M. E. Cates, Proceedings of `Dry Granular Matter', held in Cargese (Sept. 1997), to appear (Springer Verlag).

\bibitem{triangular} J.P. Bouchaud, M. E. Cates, preprint cond-mat/9801132,
submitted to `Granular Matter'.

\bibitem{lawe} W. Landry, B.T. Werner: Physica {\bf D77} (1994), 238-260

\bibitem{usinprep} O. Terzidis, Ph. Claudin, F. Rioual, A. Valance and
J.P. Bouchaud, in preparation.


\end{thebibliography}
\end{document}